\documentclass[a4,11pt]{article}

\usepackage{setspace}
\usepackage{amssymb}

\begin{document}

\pagenumbering{roman}
\newtheorem{theorem}{\sc\bf Theorem}
\newtheorem{lemma}[theorem]{\sc\bf Lemma}
\newtheorem{corollary}[theorem]{\sc\bf Corollary}
\newtheorem{proposition}[theorem]{\sc\bf Proposition}
\newcommand{\boldgreek}[1]{\mbox{\boldmath $#1$}}

\vspace{3.0cm}

\noindent
{\large\bf On a Gel'fand-Yaglom-Peres theorem for $f$-divergences}\\

\noindent
{\small\sc Gustavo L.\ Gilardoni}\footnote{
\begin{minipage}[t]{0.95\textwidth}
{\bf Address for correspondence}:
Departamento de Estat\'{\i}stica, Universidade de Bras\'{\i}lia,
Bras\'{\i}lia, DF 70910--900, Brazil. {\em e-mail}:
gilardon@unb.br. Research partially supported by CAPES, CNPq and
FINATEC grants.
\end{minipage}} \\
{\em Universidade de Bras\'{\i}lia}

\mbox{}

\noindent
\begin{minipage}[t]{0.9\textwidth}
{\small {\bf Abstract}.
It is shown that the $f$-divergence between two probability measures
$P$ and $R$ equals the supremum of the same $f$-divergence computed
over all finite measurable partitions of the original space,
thus generalizing results previously proved by
Gel'fand and Yaglom and by Peres for the Information Divergence
and more recently by
Dukkipati, Bhatnagar and Murty for the Tsallis' and R\'{e}nyi's divergences.
}
\end{minipage}

\mbox{}

\noindent
\begin{minipage}[t]{0.9\textwidth}
{\small {\bf Keywords}.
Information divergence,
Kullback-Leibler divergence,
Hellinger's discrimination,
Relative Entropy,
R\'{e}nyi's divergences,
Tsallis' divergences
}
\end{minipage}

\mbox{}

\noindent
\begin{minipage}[t]{0.9\textwidth}
{\small {\bf MSC (2000)}. 
94A17, 26D15.}
\end{minipage}

\newpage

\pagenumbering{arabic}

\section*{\normalsize 1. Introduction}
The purpose of this short note is to generalize for
arbitrary $f$-divergences a result proved by
Gel'fand and Yaglom
\cite{Gelfand-Yaglom59}
and Peres
\cite{Peres59} for the Information Divergence and,
more recently, by Dukkipati et al
\cite{Dukkipatietal07}
for Tsallis' and R\'{e}nyi's divergences.
Our method focuses on the fundamental
notion of convexity of the generating function $f$
together with some standard
integration results,
thus stressing the fact that many properties
of the Information Divergence can be extended to the general class
of $f$-divergences (cf.\
\cite{Gilardoni06vajdashort,gilardoni-pinsker10}).

The rest of the note is organized as follows.
In this introduction we set up the basic
definitions and notation and state the main result,
which is then proved in Section 2.

Consider two probability measures $P$ and $R$ on a measurable
space $(X, \mathcal{A})$ and let $p$ and $r$ be their
Radon-Nykodim derivatives with respect to a common dominating
measure $\mu$, which without loss of generality can be taken $\mu = P + Q$. The differential version of
the Information or Kullback-Leibler divergence is $I(P||R) =
\int_X \ln (p/r) \, p \, d \mu$. Gel'fand and Yaglom
\cite{Gelfand-Yaglom59}
and Perez
\cite{Peres59}
(see also
\cite[Theorem 2.4.2]{Pinsker64})
showed that
\[
I(P||R) = \sup_{\pi} \sum_{k=1}^m P(E_k) \, \ln \frac{P(E_k)}{R(E_k)} \,,
\]
where the supremum is taken over all finite measurable partitions
$\pi = \{ E_1 , \ldots , E_m \}$ ($m \geq 1$) of $X$.
In other words, by discretizing both $P$ and $R$ and computing the corresponding
divergence one can get as close as wanted to $I(P||R)$.
Recently Dukkipati et al.
\cite{Dukkipatietal07}
proved a similar result for
the R\'{e}nyi's family of divergences
\[
I_{\alpha} (P||R) = \frac{1}{\alpha -1} \, \ln
\int_X ( p/r )^{\alpha -1} \, p \, d \mu
\]
and hence also for the Tsallis' divergences
\[
T_\alpha (P||R) = \frac{1}{\alpha -1} \,
\left[ \int_X ( p/r )^{\alpha-1} \, p \, d \mu -1 \right] =
\frac{1}{\alpha-1} \left[
\exp \{ (\alpha -1) \, I_\alpha (P||R) \} -1 \right]
\]
($\alpha>0$).
Their proof rely on measure theoretic considerations along
with the inequality
$[P(E)]^\alpha \leq [ \int_E (dP/dR)^\alpha \, dR ] \, [R(E)]^{\alpha -1}$,
which follows from H\"{o}lder's Inequality.

Shortly, The $f$-divergence generated
by $f$ is $D_f (P,R) = \int f (p/r) \, r \, d \mu$,
where $f: [0,\infty) \rightarrow \mbox{\boldmath$R$}$
is convex, $f(1) =0$ and, to avoid undefined expressions,
$f(0) = \lim_{u \downarrow 0} f(u)$,
$0 \cdot f(0/0) = 0$ and
$0 \cdot f(a/0) = \lim_{\epsilon \downarrow 0} \epsilon f (a / \epsilon) =
a \lim_{u \to \infty} f(u)/u$.
The class of $f$-divergences was introduced by
Csisz\'{a}r \cite{Csis63,Csiszar67} and Ali and Silvey \cite{Ali66}
and includes, besides the Information Divergence
$I (P||R) = D_{u \, \ln u} (P,R)$ and
the family of Tsallis' divergences
$T_\alpha (P||R) = D_{[ u^\alpha -1]/(\alpha-1)} (P,R)$,
the variational distance ($f(u) = |u-1|$),
the $\chi^2$ divergence ($f(u) = (u-1)^2$),
the Hellinger discrimination ($f(u) = ( \sqrt{u} -1)^2$) and
many other distances and discrepancy measures between probability measures.
While R\'{e}nyi's divergences are not properly an $f$-divergence,
they are functions of them (i.e.\
$I_\alpha (P||Q) = (\alpha -1)^{-1} \ln [ 1 + (\alpha -1) \, T_\alpha (P||R)]$).

Our main result, of which the case of the Information and the
Tsallis' divergences are special cases, is the following.

\noindent
{\bf Proposition 1}.
{\em Let $f$ and $D_f$ be as defined above. Then for any $P$ and $R$
\begin{equation}
\label{eqn.main.df}
D_f (P,R) =
\sup_{\pi} \sum_{k=1}^m R(E_k) \, f \left( \frac{P(E_k)}{R(E_k)} \right) \,,
\end{equation}
where the supremum is taken over all finite measurable partitions $\pi$ of $X$.
} \hfill \raisebox{0.2cm}{\framebox{}}

Since the R\'{e}nyi's divergences
$I_\alpha (P||R) = (\alpha -1)^{-1} \ln [1 + (\alpha -1) T_{\alpha} (P||R)]$
are a continuous monotone function of Tsallis' divergences, it
follows from (\ref{eqn.main.df}) that
\[
I_{\alpha} (P||R) = \sup_\pi \frac{1}{\alpha -1} \, \ln \,
\sum_{k=1}^m  \frac{P(E_k)^\alpha}{R(E_k)^{\alpha -1}} \,,
\]
which is Dukkipati et al.
\cite{Dukkipatietal07} main result.

\section*{\normalsize 2. Proof of Proposition 1}
We begin with some preliminary considerations.
First, note that both sides of (\ref{eqn.main.df})
remain the same if we substitute $f(u)$ by
$\tilde{f} (u) = f(u) - a (u-1)$.
By taking $y = a(u-1)$ to be a support line to the
graph of $f$ at $u=1$
we see that we can
assume without loss of generality that $f(u)$ is nonnegative,
non increasing for $u<1$ and nondecreasing for $u>1$.
Second, since $P(A) = \int_A (p/r) \, r \, d \mu$,
if $a \leq p(x)/r(x) \leq b$ on $A$, then also $a \leq P(A)/R(A) \leq b$.
Finally, the left hand side of (\ref{eqn.main.df})
is greater than or equal than the right hand side because,
if $\pi = \{ E_j : j \in J \}$ is a finite partition of $X$,
Jensen's inequality implies that
\begin{eqnarray}
\lefteqn{
\int f(p/r) r \, d \mu =
\sum_{j \in J} R(E_j) \, \int_{E_j} f(p/r) \, \frac{r}{R(E_j)} \, d \mu } \nonumber \\
& & \geq
\sum_{j \in J} R(E_j) \, f \left( \int_{E_j} (p/r) \, \frac{r}{R(E_j)} \, d \mu \right) =
\sum_{j \in J} f \left( \frac{P(E_j)}{R(E_j)} \right) \, R(E_j) \,.
\label{eqn.referenciar}
\end{eqnarray}

We will now prove (\ref{eqn.main.df}) in the case that $D_f (P,R) < \infty$.
Due to the last consideration above, it will be enough to prove
that that the left hand side of (\ref{eqn.main.df}) is
less than or equal than the right hand side or, equivalently,
that given any $\epsilon > 0$
there exists a partition $\pi$ such that the difference between the
leftmost and the rightmost sides of (\ref{eqn.referenciar}) is less than or equal than $\epsilon$.
To do this, consider $0 < H < K$ and define
$A_H = \{ x \in X : p(x) < H \, r(x) \}$, $C_K = \{ x \in X : p(x) > K  \, r(x) \}$ and
$B_{H,K} = X - (A_H \cup C_K) = \{ x \in X : H \, r(x) \leq p(x) \leq K \, r(x) \}$.
Since $D_f (P,R) < \infty$, $\int_{A_H} f(p/r) \, r \, d \mu$ must also be finite.
Hence (i) $\lim_{H \to 0} \int_{A_H} f(p/r) \, r \, d \mu = 0$ by
dominated convergence and
(ii) also $\lim_{H \to 0} f[P(A_H)/R(A_H)] \, R(A_H) = 0$ because
Jensen's inequality implies that
$0 \leq f[P(A_H)/R(A_H)] \, R(A_H) \leq \int_{A_H} f(p/r) \, r \, d \mu$.
Therefore, for $H_0$ small enough,
$\int_{A_{H_0}} f(p/r) \, r \, d \mu - f[P(A_{H_0})/R(A_{H_0})] \, R(A_{H_0}) < \epsilon/3$.
A similar argument shows that, for $K_0$ large enough,
$\int_{C_{K_0}} f(p/r) \, r \, d \mu - f[P(C_{K_0})/R(C_{K_0})] \, R(C_{K_0}) < \epsilon/3$.
Next, since $f$ is convex, it is continuous and hence absolutely continuous in $[H_0 , K_0 ]$.
Therefore, there exists a $\delta > 0$ such that $|f(u) - f(u')| < \epsilon /3$ whenever $|u - u'| < \delta$.
With this in mind, partition the interval $[H_0 , K_0 ]$ in (say) $m$ subintervals
$I_1 , \ldots , I_m$, each having length less than $\delta$, and define
$E_i = \{ x \in X : p(x)/r(x) \in I_i \}$.
Since for $x \in E_i$ we have that $p(x)/r(x) \in I_i$ and hence also $P(E_i)/R(E_i) \in I_i$,
\begin{eqnarray*}
\lefteqn{
0 \leq \int_{E_i} f(p/r) \, r \, d \mu - f \left( \frac{P(E_i)}{R(E_i)} \right) \, R(E_i)} \\
& & =
\int_{E_i} \left\{ f(p/r) - f \left( \frac{P(E_i)}{R(E_i)} \right) \right\} \, r \, d \mu \leq
\int_{E_i} (\epsilon/3) \, r \, d \mu \leq (\epsilon / 3) \, R(E_i) \,.
\end{eqnarray*}
To finish this part of the proof, consider the partition
$\pi = \{ E_0 = A_{H_0},E_1, \ldots , E_m, E_{m+1} = C_{K_0} \}$.
The previous considerations imply that the difference between the
leftmost and the rightmost terms in (\ref{eqn.referenciar}) is less than or equal than
$\epsilon/3 + (\epsilon/3) R(B_{H_0 , K_0}) + \epsilon/3 \leq \epsilon$.

Now suppose that $D_f (P,R) = \infty$.
Then either $\int_{\{p > r\}} f(p/r) \, r \, d \mu$ or
$\int_{\{p < r\}} f(p/r) \, r \, d \mu$ should be infinite.
Suppose first that $\int_{\{p > r\}} f(p/r) \, r \, d \mu = \infty$.
We will show that there is a sequence of disjoint subsets $D_n$
such that $\sum_{n=1}^\infty f [ P(D_n) / R(D_n) ] \, R(D_n) = \infty$.
This would imply, of course, that the sets $D_n$ can be used to construct a partition of $X$
so that the rightmost term of (\ref{eqn.referenciar}) is as large as wanted,
and this in turn that the right hand side of (\ref{eqn.main.df}) is infinite.
Indeed, let $D_n = \{ x \in X \!\!: p(x) > r(x) \,\, \mbox{and} \,\, (n-1) \leq f[p(x) / r(x)] < n \}$
and for $n \geq 1$ define $b_n = \inf \{ u \in \mathbb{R} : \, u> 1 \,\, \mbox{and} \, f(u) \geq n \}$.
Since $f$ is continuous and (we are assuming wlog) nondecreasing for $u > 1$,
it follows that
$D_n = \{ x \in X \!\!: p(x) > r(x) \,\, \mbox{and} \,\, b_{n-1} \leq p(x) / r(x) < b_n \}$.
Hence, $b_{n-1} \leq P(D_n) / R(D_n) < b_n$, $(n-1) \leq f[P(D_n) / R(D_n)] < n$
and $\{f(p/r) - f[P(D_n)/R(D_n)] \} \leq 1$ in $D_n$.
Therefore,
\begin{eqnarray*}
\lefteqn{\int_{\{p > r\}} f(p/r) r \, d \mu =
\sum_{n=1}^\infty f \left( \frac{P(D_n)}{R(D_n)} \right)  R(D_n) +
\sum_{n=1}^\infty \int_{D_n} \left[ f(p/r) - f \left( \frac{P(D_n)}{R(D_n)} \right) \right] \, r \, d \mu } \\
& & \leq
\sum_{n=1}^\infty f \left( \frac{P(D_n)}{R(D_n)} \right) R(D_n) +
\sum_{n=1}^\infty \int_{D_n} r \, d \mu
\leq
\sum_{n=1}^\infty f \left( \frac{P(D_n)}{R(D_n)} \right) R(D_n) + 1 \,.
\end{eqnarray*}
This shows that if $\int_{\{p > r\}} f(p/r) \, r \, d \mu = \infty$,
then so should be $\sum_{n=1}^\infty f \left( \frac{P(D_n)}{R(D_n)} \right) R(D_n)$.
The case that $\int_{\{p < r\}} f(p/r) \, r \, d \mu = \infty$ is dealt with
in a similar manner.
\hfill \framebox{}


\begin{thebibliography}{1}

\bibitem{Gelfand-Yaglom59}
S.~I. Gel'fand and A.~M. Yaglom, ``Calculation of the amount of information
  about a random function contained in another such function,'' {\em Usp. Mat.
  Nauk.}, vol.~12, no.~1, pp.~3--52, 1959.
\newblock English translation in American Mathematical Society Translations,
  Series 2, vol. 12.

\bibitem{Peres59}
A.~Peres, ``Information {T}heory with an abstract alphabet (generalized forms
  of {M}c{M}illan's limit theorem for the case of discrete and continuous
  time),'' {\em Theory of Probability and its Applications}, vol.~4, no.~1,
  pp.~99--102, 1959.

\bibitem{Dukkipatietal07}
S.~B. A.~Dukkipati and M.~N. Murty, ``{G}elfand-{Y}aglom-{P}eres theorem for
  generalized relative entropy functionals,'' {\em Information Sciences},
  vol.~177, pp.~5707--5714, 2007.

\bibitem{Gilardoni06vajdashort}
G.~L. Gilardoni, ``On the minimum $f$-divergence for given total variation,''
  {\em C. R. Acad. Sci. Paris, Ser. I}, vol.~343, pp.~763--766, 2006.
\newblock doi:10.1016/j.crma2006.10.027.

\bibitem{gilardoni-pinsker10}
G.~L. Gilardoni, ``On {P}insker's and {V}ajda's type inequalities for
  {C}sisz{\'a}r's $f$-divergences.'' to appear in IEEE {\it Trans. Inf.
  Theory}, 2010.

\bibitem{Pinsker64}
M.~S. Pinsker, {\em Information and Information Stability of Random Variables
  and Processes}.
\newblock A. Feinstein, tr. and ed., San Francisco: Holden-Day, 1964.

\bibitem{Csis63}
I.~Csisz{\'a}r, ``Eine informationstheoretische {U}ngleichung und ihre
  anwendung auf den {B}eweis der ergodizit{\"a}t von {M}arkoffschen {K}etten,''
  {\em Publ. Math. Inst. Hungar. Acad.}, vol.~8, pp.~95--108, 1963.

\bibitem{Csiszar67}
I.~Csisz{\'a}r, ``Information-type measures of difference of probability
  distributions and indirect observations,'' {\em Studia Sci. Math. Hungar.},
  vol.~2, pp.~299--318, 1967.

\bibitem{Ali66}
S.~M. Ali and S.~D. Silvey, ``A general class of coefficients of divergence of
  one distribution from another,'' {\em J. Roy. Statist. Soc. Ser B}, vol.~28,
  pp.~131--142, 1966.

\end{thebibliography}

\bibliographystyle{ieeetr}

\end{document}